\documentstyle[12pt]{article}
\newlength{\extraspace}
\newlength{\extraspaces}
\newcommand{\be}{\begin{equation}
\addtolength{\abovedisplayskip}{\extraspaces}
\addtolength{\belowdisplayskip}{\extraspaces}
\addtolength{\abovedisplayshortskip}{\extraspace}
\addtolength{\belowdisplayshortskip}{\extraspace}}
\newcommand{\ee}{\end{equation}}
\newcommand{\ba}{\begin{eqnarray}
\addtolength{\abovedisplayskip}{\extraspaces}
\addtolength{\belowdisplayskip}{\extraspaces}
\addtolength{\abovedisplayshortskip}{\extraspace}
\addtolength{\belowdisplayshortskip}{\extraspace}}
\newcommand{\ea}{\end{eqnarray}}
\newcommand{\nonu}{\nonumber \\[.5mm]}
\newcommand{\A}{&\!\!\!}
\begin{document}
\thispagestyle{empty}
\setlength{\baselineskip}{6mm}
\begin{flushright}
SIT-LP-10/06 \\
June, 2010
\end{flushright}
\vspace{7mm}
\begin{center}
{\large \bf On SUSY breaking from NL/L SUSY relation
} \\[20mm]
{\sc Kazunari Shima}
\footnote{
\tt e-mail: shima@sit.ac.jp} \ 
and \ 
{\sc Motomu Tsuda}
\footnote{
\tt e-mail: tsuda@sit.ac.jp} 
\\[5mm]
{\it Laboratory of Physics, 
Saitama Institute of Technology \\
Fukaya, Saitama 369-0293, Japan} \\[20mm]
\begin{abstract}
We show in two dimensional space-time ($d = 2$) the relation between a $N = 2$ nonlinear supersymmetric (NLSUSY) model 
and a $N = 2$ linear (L) SUSY Yang-Mills (SYM) theory with matter ($N = 2$ LSUSY QCD theory). 
We give a new interpretation of four Nambu-Goldstone fermion (superon) contact terms, 
which emerge from a $N = 2$ general SUSY QCD (composite) action, as mass terms for LSUSY supermultiplets 
and discuss the possible SUSY breaking mechanism in NL/L SUSY relation for SUSY gauge theories in $d = 2$. 
\\[5mm]
\noindent
PACS: 11.30.Pb, 12.60.Jv, 12.60.Rc, 12.10.-g \\[2mm]
\noindent
Keywords: supersymmetry, Nambu-Goldstone fermion, 
nonlinear/linear SUSY relation, composite unified theory 
%
%
\end{abstract}
\end{center}

\newpage

\noindent
Dynamics of massless Nambu-Goldstone (NG) fermions for nonlinear (NL) representation \cite{VA} 
of supersymmetry (SUSY) \cite{GL}-\cite{WZ2} is described in the NLSUSY theory \cite{VA} 
where spontaneous SUSY breaking mechanism is encoded a priori. 
NLSUSY general relativity theory (NLSUSY GR) \cite{KS1} for SGM scenario \cite{KS1}-\cite{ST3-SGM} 
gives in the asymptotic Riemann-flat space-time the (fundamental) NLSUSY model 
with a fixed dimensional constant (the SUSY breaking scale) 
depending on the cosmological and the gravitational constants. 

The low energy effective theory of NLSUSY GR is obtained by means of the linearization of NLSUSY \cite{IK1}-\cite{UZ}, 
which gives the relation between the NLSUSY model and linear (L) SUSY theories (with spontaneous SUSY breaking) 
(abbreviated as NL/L SUSY relation) \cite{IK1}-\cite{ST2-lin} 
including the relation for Yukawa interactions \cite{ST3-lin,ST4-lin}, SUSY QED \cite{ST5-lin,ST6-lin} 
and super Yang-Mills (SYM) theories \cite{ST7-lin}, etc. 
In NL/L SUSY relation based on NLSUSY GR, LSUSY-supermultiplet fields are realized 
as the composite (massless) eigenstates of the NG fermions ({\it superons}) on true vacuum, 
where the large scale structure of space-time is related to low energy particle physics. 
Indeed, the NL/L SUSY relation for $N = 2$ SUSY QED theory 
(in two-dimensional space-time ($d = 2$) for simplicity of calculations) \cite{ST5-lin,ST6-lin} 
gives, e.g. a natural explanation of the mysterious (observed) numerical relation 
between the (four dimensional) dark energy density of the universe and the neutrino ($\nu$) mass \cite{ST3-SGM,ST-vac,STL}, 
provided $\nu$ is a composite of superons of these kinds. 
(Note that for $N = 2$ SUSY in SGM scenario $J^P = 1^-$ gauge field appears \cite{STT2} in the linearization of NLSUSY. 
Therefore $N = 2$ SUSY is realistic minimal case.) 

In order to extend the abovementioned analysis of low energy particle physics to non-Abelian gauge theories, 
it is important to study further the SUSY breaking mechanism for SUSY gauge theories in NL/L SUSY relation. 
In this letter we show in $d = 2$ that a $N = 2$ LSUSY SYM theory with matter (a $N = 2$ LSUSY QCD theory) 
is related to the $N = 2$ NLSUSY model by expressing basic fields in LSUSY QCD in terms of NG-fermion superons. 
We focus on the $d = 2$, $N = 2$ SUSY QCD {\it composite} theory 
and discuss adequate SUSY breaking terms in NL/L SUSY relation. 
We shall see in the relation between the NLSUSY action and a general SUSY QCD action 
that four superon contact terms emerge from the SUSY QCD action 
for redundant auxiliary fields (except $D$ and $F$ auxiliary fields) 
as the similar case for the SUSY QED theory \cite{ST5-lin,ST6-lin}. 
By considering a new interpretation of those four superon contact terms as mass terms for LSUSY supermultiplets, 
the possible SUSY breaking mechanism in NL/L SUSY relation for SUSY gauge theories is discussed in $d = 2$. 

The fundamental action of the NLSUSY model \cite{VA,BV} for $N$ SUSY 
in terms of (Majorana) superons $\psi^i(x)$ ($i, j, \cdots = 1, \cdots, N$) is given by 
\be
S_{\rm NLSUSY} = - {1 \over {2 \kappa^2}} \int d^2x \ \vert w \vert, 
\label{NLSUSYaction}
\ee
where $\kappa$ is a dimensional constant whose dimension is (mass)$^{-1}$ in $d = 2$ 
\footnote{
Minkowski space-time indices in $d = 2$ are denoted by $a, b, \cdots = 0, 1$. 
The Minkowski space-time metric is 
${1 \over 2}\{ \gamma^a, \gamma^b \} = \eta^{ab} = {\rm diag}(+, -)$ 
and $\sigma^{ab} = {i \over 2}[\gamma^a, \gamma^b] 
= i \epsilon^{ab} \gamma_5$ $(\epsilon^{01} = 1 = - \epsilon_{01})$, 
where we use the $\gamma$ matrices defined as $\gamma^0 = \sigma^2$, 
$\gamma^1 = i \sigma^1$, $\gamma_5 = \gamma^0 \gamma^1 = \sigma^3$ 
with $(\sigma^1, \sigma^2, \sigma^3)$ being Pauli matrices. 
}
and the determinant $\vert w \vert$ is defined as 
\be
\vert w \vert = \det(w^a{}_b) = \det(\delta^a_b + t^a{}_b) 
\ee
with $t^a{}_b = - i \kappa^2 \bar\psi^i \gamma^a \partial_b \psi^i$, 
which terminates at ${\cal O}(t^2)$ for the $d = 2$ case. 
The $N$ NLSUSY action (\ref{NLSUSYaction}) is invariant under the following NLSUSY transformations of $\psi^i$, 
\be
\delta_\zeta \psi^i = {1 \over \kappa} \zeta^i 
- i \kappa \bar\zeta^j \gamma^a \psi^j \partial_a \psi^i, 
\label{NLSUSY}
\ee
which are parametrized by means of constant (Majorana) spinor parameters $\zeta^i$. 
The NLSUSY transformations (\ref{NLSUSY}) satisfy a closed off-shell commutator algebra, 
\be
[\delta_{\zeta_1}, \delta_{\zeta_2}] = \delta_P(\Xi^a), 
\label{NLSUSYcomm}
\ee
where $\delta_P(\Xi^a)$ means a translation with the parameters 
$\Xi^a = 2 i \bar\zeta_1^i \gamma^a \zeta_2^i$. 

In NL/L SUSY relation, the relations between basic fields in LSUSY theories and superons $\psi^i$ 
(called {\it SUSY invariant relations}) are obtained in the superfield formulation \cite{WB}. 
They are derived systematically from the hybrid transformations of L and NLSUSY 
on specific supertranslations of superspace coordinates $\{ x^a, \theta_\alpha^i \}$ parametrized by $\psi^i$ \cite{IK1}, 
\ba
\A \A 
x'^a = x^a + i \kappa \bar\theta^i \gamma^a \psi^i, 
\nonu
\A \A 
\theta'^i = \theta^i - \kappa \psi^i, 
\label{specific}
\ea
and the adoption of the subsequent {\it SUSY invariant constraints} for superfields. 
(As for the detailed prescription for the construction of the SUSY invariant relations 
in the superfield formulation, see \cite{IK1,UZ}.)

Let us show below SUSY invariant relations for $N = 2$ vector and $N = 2$ scalar matter supermultiplets 
in $d = 2$, $N = 2$ SUSY QCD theory. 
The $N = 2$ general gauge \cite{DVF,ST-SF} and the $N = 2$ scalar superfields 
on $N = 2$ superspace ($i = 1, 2$) are defined in $d = 2$ as 
\ba
{\cal V}(x, \theta) \A = \A C(x) + \bar\theta^i \Lambda^i(x) 
+ {1 \over 2} \bar\theta^i \theta^j M^{ij}(x) 
- {1 \over 2} \bar\theta^i \theta^i M^{jj}(x) 
+ {1 \over 4} \epsilon^{ij} \bar\theta^i \gamma_5 \theta^j \phi(x) 
\nonu
\A \A 
- {i \over 4} \epsilon^{ij} \bar\theta^i \gamma_a \theta^j v^a(x) 
- {1 \over 2} \bar\theta^i \theta^i \bar\theta^j \lambda^j(x) 
- {1 \over 8} \bar\theta^i \theta^i \bar\theta^j \theta^j D(x), 
\label{VSF}
\\
\Phi^{iA}(x, \theta) \A = \A B^{iA}(x) + \bar\theta^i \chi^A(x) - \epsilon^{ij} \bar\theta^j \nu^A(x) 
- {1 \over 2} \bar\theta^j \theta^j F^{iA}(x) + \bar\theta^i \theta^j F^{jA}(x) 
- i \bar\theta^i \!\!\not\!\partial B^{jA}(x) \theta^j 
\nonu
\A \A 
+ {i \over 2} \bar\theta^j \theta^j (\bar\theta^i \!\!\not\!\partial \chi^A(x) 
- \epsilon^{ik} \bar\theta^k \!\!\not\!\partial \nu^A(x)) 
+ {1 \over 8} \bar\theta^j \theta^j \bar\theta^k \theta^k \Box B^{iA}(x). 
\label{SSF}
\ea
In the gauge superfield (\ref{VSF}) we denote $C, D$ and $M^{ij} = M^{(ij)}$ 
$\left(= {1 \over 2}(M^{ij} + M^{ji}) \right)$ for scalar fields, 
$\Lambda^i$ and $\lambda^i$ for (Majorana) spinor fields, $\phi$ for pseudo scalar fields 
and $v^a$ for vector fields, respectively. 
These component fields $V(x)$ in Eq.(\ref{VSF}) belong to the adjoint representation of gauge group $G$; 
namely, $V(x) = V^I(x) T^I$ with generators $T^I = (T^I)^A{}_B$ of $G$ satisfying $[T^I, T^J] = i f^{IJK} T^K$. 
In the scalar superfields (\ref{SSF}) we denote $B^{iA}$ for scalar fields, 
$\chi^A$ and $\nu^A$ for (Majorana) spinor fields and $F^{iA}$ for auxiliary scalar fields. 

Note that the $N = 2$ minimal off-shell vector supermultiplet are defined 
from the general component fields in Eq.(\ref{VSF}) by 
\be
\{ v^{aI}_0, \lambda^{iI}_0, A^I_0, \phi^I_0, D^I_0 \} 
= \{ v^{aI}, \lambda^{iI} + i \!\!\not\!\partial \Lambda^{iI}, M^{iiI}, \phi^I, D^I + \Box C^I \}, 
\label{minimal}
\ee
whose LSUSY transformations are apparently expressed in terms of only the fields (\ref{minimal}) 
except a gauge parameter composed of $\Lambda^{iI}$ 
and satisfy the commutator algebra (\ref{NLSUSYcomm}) (for example, see \cite{ST4-lin}). 

By extending the previous works \cite{ST2-lin} for NL/L SUSY relation in the superfield formulation 
to the superfields (\ref{VSF}) and (\ref{SSF}), 
SUSY invariant relations for the above component fields in SUSY QCD 
are given as the composites of $\psi^i$ for the $N = 2$ vector supermultiplet, 
\ba
C^I \A = \A - {1 \over 8} \xi^I \kappa^3 \bar\psi^i \psi^i \bar\psi^j \psi^j \vert w \vert, 
\nonu
\Lambda^{iI} \A = \A - {1 \over 2} \xi^I \kappa^2 
\psi^i \bar\psi^j \psi^j \vert w \vert, 
\nonu
M^{ijI} \A = \A {1 \over 2} \xi^I \kappa \bar\psi^i \psi^j \vert w \vert, 
\nonu
\phi^I \A = \A - {1 \over 2} \xi^I \kappa \epsilon^{ij} \bar\psi^i \gamma_5 \psi^j \vert w \vert, 
%
\nonu
v^{aI} \A = \A - {i \over 2} \xi^I \kappa \epsilon^{ij} \bar\psi^i \gamma^a \psi^j \vert w \vert, 
\nonu
\lambda^{iI} \A = \A \xi^I \psi^i \vert w \vert, 
\nonu
D^I \A = \A {\xi^I \over \kappa} \vert w \vert, 
\label{VSUSYinv}
\end{eqnarray}
and for the $N = 2$ scalar matter supermultiplet, 
\ba
\chi^A \A = \A \xi^{iA} \left[ \psi^i \vert w \vert
+ {i \over 2} \kappa^2 \partial_a 
( \gamma^a \psi^i \bar\psi^j \psi^j \vert w \vert 
) \right], 
\nonu
B^{iA} \A = \A - \kappa \left( {1 \over 2} \xi^{iA} \bar\psi^j \psi^j 
- \xi^{jA} \bar\psi^i \psi^j \right) \vert w \vert, 
\nonu
\nu^A \A = \A \xi^{iA} \epsilon^{ij} \left[ \psi^j \vert w \vert 
+ {i \over 2} \kappa^2 \partial_a 
( \gamma^a \psi^j \bar\psi^k \psi^k \vert w \vert 
) \right], 
\nonu
F^{iA} \A = \A {\xi^{iA} \over \kappa} \left\{ \vert w \vert 
+ {1 \over 8} \kappa^3 
\Box ( \bar\psi^j \psi^j \bar\psi^k \psi^k \vert w \vert ) 
\right\} 
\nonu
\A \A 
- i \kappa \xi^{jA} \partial_a ( \bar\psi^i \gamma^a \psi^j \vert w \vert ), 
\label{SSUSYinv}
\end{eqnarray}
which are the form containing some vanishing terms due to $(\psi^i)^5 \equiv 0$. 
In the SUSY invariant relations (\ref{VSUSYinv}) and (\ref{SSUSYinv}), 
$\xi^I$ and $\xi^{iA}$ with the indices $I$ and $A$ for the gauge group $G$ 
are arbitrary real constants corresponding to the constant terms of the auxiliary fields 
$D^I = D^I(\psi)$ and $F^{iA} = F^{iA}(\psi)$ according to the {\it simplest} SUSY invariant constraints 
\cite{ST2-lin,ST6-lin} (see also \cite{ST-gauge} as for more general SUSY invariant constraints). 

Here we introduce a LSUSY action for $d = 2$, $N = 2$ SUSY QCD theory in order to discuss NL/L SUSY relation for SUSY gauge theories. 
The SUSY QCD (gauge invariant) action is the sum of a pure SYM action for the vector supermultiplet 
and a matter action for the scalar supermultiplet coupled to the vector one, i.e. 
\be
S_{\rm SQCD} = S_{\rm SYM} + S_{\rm matter}. 
\label{SQCD}
\ee
The $N = 2$ LSUSY SYM action $S_{\rm SYM}$ in $d = 2$ is given in terms of the component fields (\ref{minimal}) by 
\ba
S_{\rm SYM} \A = \A \int dx^2 \ {\rm tr} 
\Big\{ -{1 \over 2} (F_{0ab})^2 + i \bar\lambda_0^i \!\!\not\!\!D \lambda_0^i + (D_a A_0)^2 + (D_a \phi_0)^2 + D_0^2 
\nonu
\A \A 
- 2ig (\epsilon^{ij} A_0 \bar\lambda_0^i \lambda_0^j + \phi_0 \bar\lambda_0^i \gamma_5 \lambda_0^i) 
+ g^2 [A_0, \phi_0]^2 \Big\} 
\label{SYM}
\ea
where $F_{0ab} = \partial_a v_{0b} - \partial_b v_{0a} - ig [v_{0a}, v_{0b}]$ 
and $D_a \varphi = \partial_a \varphi - ig [v_{0a}, \varphi]$ (for the vector supermultiplet components $\varphi$) 
with a gauge coupling constant $g$. 

The LSUSY matter action $S_{\rm matter}$ is defined from the gauge and the scalar superfields (\ref{VSF}) and (\ref{SSF}) as 
\be
S_{\rm matter} = - {1 \over 8} \int d^2 x \int d^4 \theta^- \ \Phi^\dagger_A (e^{-4g{\cal V}})^A{}_B \Phi^B 
\label{matter-gen}
\ee
where $\Phi^A(x, \theta) = {1 \over \sqrt{2}} \{ \Phi^{1A}(x, \theta) + i \Phi^{2A}(x, \theta) \}$ 
and $\theta^{\pm} = \theta^1 \pm i \theta^2$. 
The action (\ref{matter-gen}) are separated into the following two parts, $S^0_{\rm matter}$ 
in terms of the fields (\ref{minimal}) for the minimal off-shell vector supermultiplet 
and $S'_{\rm matter}$ in terms of the fields containing explicitly the redundant component fields 
$\{ C, \Lambda^i, M^{12}, M^{11} - M^{22} \}$, i.e. 
\be
S_{\rm matter} = S^0_{\rm matter} + S'_{\rm matter}. 
\ee
Indeed, the actions $S^0_{\rm matter}$ and $S'_{\rm matter}$ are given by 
\ba
S^0_{\rm matter} 
\A = \A \int d^2 x \ \Big[ \ i \bar\Psi_A \!\!\not\!\!D \Psi^A 
+ \vert D_a B^A \vert^2 + \vert F^A \vert^2 
\nonu
\A \A 
+ g [ \sqrt{2} \{ \bar\Psi_A (\lambda_0)^A{}_B B^B + B^*_A (\bar\lambda_0)^A{}_B \Psi^B \} 
- B^*_A (D_0)^A{}_B B^B 
\nonu
\A \A 
+ \bar\Psi_A (A_0)^A{}_B \Psi^B + i \bar\Psi_A \gamma_5 (\phi_0)^A{}_B \Psi^B ] 
\nonu
\A \A 
- g^2 B^*_A \{ (A_0 A_0)^A{}_B + (\phi_0 \ \phi_0)^A{}_B \} B^B \ \Big], 
\label{matter-min}
\\[2mm]
S'_{\rm matter} 
\A = \A \int d^2 x \ g \ [ \ 2i \{ F^*_A (M_{12})^A{}_B B^B - B^*_A (M_{12})^A{}_B F^B \} 
\nonu
\A \A 
- F^*_A (M_{11} - M_{22})^A{}_B B^B - B^*_A (M_{11} - M_{22})^A{}_B F^B 
\nonu
\A \A 
+ 2 \sqrt{2} \{ \bar\Psi_A (\Lambda^*)^A{}_B F^B + F^*_A (\bar\Lambda^*)^A{}_B \Psi^B \} 
\nonu
\A \A 
- 4 F^*_A (C)^A{}_B F^B 
+ \cdots] + \cdots, 
\label{matter-red}
\ea
where we define the (complex) component fields $(B^A, \Psi^A, F^A)$ 
from $(B^{iA}, \chi^A, \nu^A, F^{iA})$ for the scalar supermultiplet 
and the spinor fields $\lambda_0$ from $\lambda_0^i$ as 
\ba
\A \A B^A = {1 \over \sqrt{2}} (B^{1A} + i B^{2A}), 
\ \ \ F^A = {1 \over \sqrt{2}} (F^{1A} - i F^{2A}), 
\nonu
\A \A 
\Psi^A = {1 \over \sqrt{2}} (\chi^A + i \nu^A), 
\ \ \ \lambda_0 = {1 \over \sqrt{2}} (\lambda_0^1 - i \lambda_0^2), 
\ea
and also $D_a \varphi^A = \partial_a \varphi^A - ig (v_{0a})^A{}_B \varphi^B$ 
and $D_a \varphi^*_A = \partial_a \varphi^*_A + ig \varphi^*_B (v_{0a})^B{}_A$ 
(for the scalar supermultiplet components $\varphi^A$). 

Now we discuss the relation between the NLSUSY action (\ref{NLSUSYaction}) for $N = 2$ SUSY and the LSUSY QCD action (\ref{SQCD}). 
By substituting the SUSY invariant relations (\ref{VSUSYinv}) for the vector supermultiplet into the SYM action (\ref{SYM}), 
it reduces to Eq.(\ref{NLSUSYaction}) as 
\be
S_{\rm SYM}(\psi) = - (\xi^I)^2 S_{\rm NLSUSY} + [{\rm surface\ terms}], 
\label{relation-SYM}
\ee
which is a trivial relation in a sense that each interaction terms in $S_{\rm SYM}(\psi)$ at ${\cal O}(g)$ and ${\cal O}(g^2)$ 
vanish due to $(\psi^i)^5 \equiv 0$ in $d = 2$. 
However, the relation (\ref{relation-SYM}) in $d = 2$ is nontrivial for $N = 3$ SUSY \cite{ST7-lin}, 
where (non-vanishing) interaction terms at ${\cal O}(g)$ in terms of $\psi^i$ cancel with each other in $S_{\rm SYM}(\psi)$. 

As for the matter action (\ref{matter-gen}), we can show the relation between $S_{\rm matter}$ and $S_{\rm NLSUSY}$ 
by changing the integration variables $\{ x^a, \theta_\alpha^i \}$ to $\{ x'^a, {\theta'}_\alpha^i \}$ 
of Eq.(\ref{specific}), 
\ba
S_{\rm matter}(\psi) 
\A = \A - {1 \over 8} \int d^2 x \int d^4 \theta^- J 
\ \tilde\Phi^\dagger_A (e^{-4g \tilde{\cal V}})^A{}_B \tilde\Phi^B 
\nonu
\A = \A - 2 \vert \xi^A \vert^2 S_{\rm NLSUSY}. 
\label{relation-matter}
\ea
In this relation (\ref{relation-matter}), 
$J = J(x,\theta)$ is the Jacobian \cite{IK1,UZ,STT1,ST2-lin} which is proportional to the determinant $\vert w \vert$ 
and $\tilde\Phi^A = \tilde\Phi^A(x,\theta)$ and $\tilde{\cal V}^I = \tilde{\cal V}^I(x,\theta)$ 
corresponding to the superfields on $\{ x'^a, {\theta'}_\alpha^i \}$ are 
\be
\tilde\Phi^A(x, \theta) = {\xi^A \over {2 \kappa}} \bar\theta^- \theta^+, 
\ \ \tilde{\cal V}^I(x, \theta) = - {\xi^I \over {8 \kappa}} \bar\theta^- \theta^- \bar\theta^- \theta^-, 
\ee
under the simplest SUSY invariant constraints, 
$\tilde D^I(x) = {\xi^I \over \kappa}$ and $\tilde F^A(x) = {\xi^A \over \kappa}$ 
(${\rm the\ other\ tilded\ component\ fields\ in}\ \tilde\Phi^A \ {\rm and}\ \tilde{\cal V}^I = 0$), 
where $\xi^A = {1 \over \sqrt{2}} (\xi^{1A} - i \xi^{2A})$. 

From Eqs.(\ref{relation-SYM}) and (\ref{relation-matter}), the LSUSY QCD action (\ref{SQCD}) is related 
to the NLSUSY action (\ref{NLSUSYaction}) as 
\be
S_{\rm SQCD}(\psi) = (S_{\rm SYM} + S_{\rm matter})(\psi) = - \{ (\xi^I)^2 + 2 \vert \xi^A \vert^2 \} S_{\rm NLSUSY}. 
\label{relation-SQCD}
\ee

Let us discuss further the NL/L SUSY relation for the minimal off-shell vector supermultiplet (\ref{minimal}). 
Substituting the SUSY invariant relations (\ref{VSUSYinv}) and (\ref{SSUSYinv}) into the actions 
(\ref{matter-min}) and (\ref{matter-red}) in $S_{\rm matter}$ produces the following four superon contact terms, 
\ba
S^0_{\rm matter}(\psi) \ {\rm at} \ {\cal O}(g) 
\A = \A \int d^2x \left\{ {1 \over 2} g \kappa \xi^*_A (\xi)^A{}_B \xi^B \bar\psi^i \psi^i \bar\psi^j \psi^j \right\}, 
\label{4fermi-min} \\[1mm]
S'_{\rm matter}(\psi) \ {\rm at} \ {\cal O}(g) 
\A = \A \int d^2x \left\{ -{1 \over 2} g \kappa \xi^*_A (\xi)^A{}_B \xi^B \bar\psi^i \psi^i \bar\psi^j \psi^j \right\}. 
\label{4fermi-red}
\ea
Namely, the general NL/L SUSY relation (\ref{relation-matter}) holds by means of the (nontrivial) cancellations 
among the terms (\ref{4fermi-min}) and (\ref{4fermi-red}) as 
\be
S_{\rm matter}(\psi) = S^0_{\rm matter}(\psi) 
+ \int d^2x \left\{ -{1 \over 2} g \kappa \xi^*_A (\xi)^A{}_B \xi^B \bar\psi^i \psi^i \bar\psi^j \psi^j \right\} 
= - 2 \vert \xi^A \vert^2 S_{\rm NLSUSY}, 
\label{relation-4fermi}
\ee
which is similar to the case of SUSY QED theory \cite{ST5-lin,ST6-lin}. 

Here let us propose a new interpretation of the four superon contanct terms (\ref{4fermi-red}) 
in the relation (\ref{relation-4fermi}) as some SUSY breaking terms. 
We identify the contact terms (\ref{4fermi-red}) with scalar mass terms responsible for soft SUSY breaking 
(for example, see \cite{CEKKLW} and references there in). The follwing $G$-invariant mass terms for the scalar fields, 
\be
S_{\rm mass} 
= \int d^2x \ [ \ - \{ (\mu^2)^A{}_B B^*_A B^B + \mu_A^2 \ {\rm tr} \ A_0^2 + \mu_\phi^2 \ {\rm tr} \ \phi_0^2 \} \ ], 
\label{mass}
\ee
are considered in the $d = 2$, $N = 2$ SUSY QCD theory. 
This becomes four superon contact terms, 
\be
S_{\rm mass}(\psi) 
= \int d^2x \ \kappa^2 \left\{ {1 \over 2} \xi^*_A (\mu^2)^A{}_B \xi^B - {1 \over 8} (\mu_A^2 + \mu_\phi^2) (\xi^I)^2 \right\} 
\bar\psi^i \psi^i \bar\psi^j \psi^j, 
\ee
under the SUSY invariant relations (\ref{VSUSYinv}) and (\ref{SSUSYinv}). 
Then the relation (\ref{relation-4fermi}) can be written as 
\be
S_{\rm matter}(\psi) = (S^0_{\rm matter} + S_{\rm mass})(\psi) = - 2 \vert \xi^A \vert^2 S_{\rm NLSUSY}, 
\label{relation-mass}
\ee
provided 
\be
\xi^*_A \left\{ (\mu^2)^A{}_B + {g \over \kappa} (\xi)^A{}_B \right\} \xi^B 
- {1 \over 4} (\mu_A^2 + \mu_\phi^2) (\xi^I)^2 = 0. 
\label{constraint-mass}
\ee

By adding the relation (\ref{relation-SYM}) to Eq.(\ref{relation-mass}), we show the NL/L SUSY relation 
for the $d = 2$, $N = 2$ (massive) SUSY QCD theory with the SUSY breaking terms, 
\be
- S_{\rm NLSUSY} = S_{\rm SQCD} = S^0_{\rm SQCD} + S_{\rm mass} + [{\rm surface\ terms}] 
\label{relation-SQCDmass}
\ee
when $(\xi^I)^2 + 2 \vert \xi^A \vert^2 = 1$. 
LSUSY in the SUSY QCD action $S^0_{\rm SQCD} = S_{\rm SYM} + S^0_{\rm matter}$ 
for the minimal off-shell vector supermultiplet is broken explicitly by means of $S_{\rm mass}$. 
Note that adding linear $D$ (or $D_0$) terms \cite{FI}, 
\be
S_D = \int d^2x \left( -{\xi^I \over \kappa} D^I \right), 
\ee
which provides the vacuum expectation values $<D^I> = {\xi^I \over \kappa}$, 
to the NL/L SUSY relation (\ref{relation-SQCDmass}) gives the relation for $+ S_{\rm NLSUSY}$ as 
\be
S_{\rm NLSUSY} = S_{\rm SQCD} + S_D = S^0_{\rm SQCD} + S_{\rm mass} + S_D + [{\rm surface\ terms}], 
\label{relation-D}
\ee
when $(\xi^I)^2 - 2 \vert \xi^A \vert^2 = 1$ which is favorable from SGM scenario. 
The above arguments also hold in NL/L SUSY relation for the $d = 2$, $N = 2$ SUSY QED theory \cite{ST5-lin,ST6-lin}. 

Our results are summarized as follows. 
In this letter we have shown in $d = 2$ that the (fundamental) $N = 2$ NLSUSY action (\ref{NLSUSYaction}) 
is related to the $N = 2$ LSUSY QCD action (\ref{SQCD}) as the NL/L SUSY relation (\ref{relation-SQCD}) 
by constructing the SUSY invariant relations (\ref{VSUSYinv}) and (\ref{SSUSYinv}) 
with the constants $\xi^I$ and $\xi^{iA}$ corresponding to the constant terms (vacuum expectation values) 
of the auxiliary fields $D^I = D^I(\psi)$ and $F^{iA} = F^{iA}(\psi)$. 
The four superon contact terms (\ref{4fermi-min}) and (\ref{4fermi-red}), 
which emerge from the actions $S^0_{\rm matter}$ and $S'_{\rm matter}$ at ${\cal O}(g)$, 
cancel with each other in the NL/L SUSY relation. 
We have pointed out another role (viewpoint) of NL/L SUSY relation, which induces LSUSY soft SUSY breaking. 
By identifying the contact terms (\ref{4fermi-red}) with the scalar mass terms (\ref{mass}), 
we have shown the new NL/L SUSY relation (\ref{relation-SQCDmass}) (or (\ref{relation-D})) 
with the constraint equation (\ref{constraint-mass}) for the mass parameters 
$(\mu^2)^A{}_B$, $\mu_A^2$ and $\mu_\phi^2$ by means of $\xi^I$ and $\xi^{iA}$, 
in which LSUSY is broken explicitly. 
It is interesting if these arguments may have some relations to the familiar explicit soft SUSY breaking 
by $A$ term in LSUSY gauge model. 
These arguments for the $d = 4$ case are interesting and open problems. 
\vspace{1cm}

\noindent
{\Large{\bf Acknowledgements}} \\[2mm]
One of the authors (K.S.) would like to thank Professor K. Shizuya and Professor T. Eguchi 
for the warm hospitalities at YITP, Kyoto University for the support during the stay.

\newpage

%
\newcommand{\NP}[1]{{\it Nucl.\ Phys.\ }{\bf #1}}
\newcommand{\PL}[1]{{\it Phys.\ Lett.\ }{\bf #1}}
\newcommand{\CMP}[1]{{\it Commun.\ Math.\ Phys.\ }{\bf #1}}
\newcommand{\MPL}[1]{{\it Mod.\ Phys.\ Lett.\ }{\bf #1}}
\newcommand{\IJMP}[1]{{\it Int.\ J. Mod.\ Phys.\ }{\bf #1}}
\newcommand{\PR}[1]{{\it Phys.\ Rev.\ }{\bf #1}}
\newcommand{\PRL}[1]{{\it Phys.\ Rev.\ Lett.\ }{\bf #1}}
\newcommand{\PTP}[1]{{\it Prog.\ Theor.\ Phys.\ }{\bf #1}}
\newcommand{\PTPS}[1]{{\it Prog.\ Theor.\ Phys.\ Suppl.\ }{\bf #1}}
\newcommand{\AP}[1]{{\it Ann.\ Phys.\ }{\bf #1}}

\end{document}